\documentclass[conference]{IEEEtran}
\IEEEoverridecommandlockouts

\usepackage{cite}
\usepackage{amsmath,amssymb,amsfonts}
\usepackage{algorithm}
\usepackage{algorithmicx}
\usepackage{algpseudocode}
\usepackage{graphicx}
\usepackage{textcomp}
\usepackage{xcolor}
\usepackage{subfig}
\usepackage[T1]{fontenc}
\usepackage[acronym]{glossaries}
\usepackage{mathtools}
\usepackage{bbold} 
\usepackage{bm} 
\usepackage{multirow}
\usepackage{booktabs} 
\usepackage{threeparttable} 
\usepackage{color} 
\usepackage{grffile}
\usepackage{xspace}
\usepackage{tikz}
\usetikzlibrary{fit,shapes.geometric,decorations.pathreplacing,arrows.meta}
\usepackage{subfig}
\algnewcommand{\OnelineIf}[1]{\State\algorithmicif\ #1\ \algorithmicthen}
\algnewcommand{\OnelineForAll}[1]{\State\algorithmicfor\ #1\ \algorithmicdo}

\def\BibTeX{{\rm B\kern-.05em{\sc i\kern-.025em b}\kern-.08em
    T\kern-.1667em\lower.7ex\hbox{E}\kern-.125emX}}

\usepackage{CJKutf8}

\renewcommand{\[}{\left[}
\renewcommand{\]}{\right]}
\renewcommand{\(}{\left(}
\renewcommand{\)}{\right)}
\DeclareMathOperator*{\argmax}{arg\,max}

\newcommand{\Hermit}{\dagger}
\newcommand{\defeq}{\coloneqq}
\newcommand{\Capacity}{C}
\newcommand{\AvgCapacity}{\bar{C}}
\newcommand{\CapBest}{C^{\mathrm{best}}}
\newcommand{\Fairness}{F}

\newcommand{\rssi}{\mathrm{RSSI}}
\newcommand{\NumRxAntenna}{n_{\mathrm{r}}}
\newcommand{\NumTxAntenna}{n_{\mathrm{t}}}
\newcommand{\NumSubcarrier}{K}
\newcommand{\CSI}{H}
\newcommand{\NormCSI}{\tilde{H}}
\newcommand{\SCindex}{k}
\newcommand{\BWperSubCarrier}{\Delta f}
\newcommand{\NoiseDensity}{N_0}
\newcommand{\IdentityMat}{I}
\newcommand{\Boltzman}{k_\mathrm{B}}
\newcommand{\Temperature}{T}
\newcommand{\NoiseFactor}{F}
\newcommand{\Stations}{\mathcal{S}}

\newcommand{\ComplexMat}{\mathbb{C}}
\newcommand{\AngleSet}{\bm{\theta}}
\newcommand{\AngleSetIdeal}{\AngleSet^*}
\newcommand{\AngleSetBest}{\AngleSet^{\mathrm{best}}}
\newcommand{\StaSet}{S}

\newcommand{\gainCoef}{\alpha}
\newcommand{\WaitTime}{T_\mathrm{wait}}
\newcommand{\AvgTime}{T_\mathrm{avg}}
\newcommand{\Opt}[1]{\mathcal{O}[#1]}
\newcommand{\OptSuggest}{\Opt{\StaSet}.\mathit{suggest}()}
\newcommand{\OptReport}[2]{\Opt{\StaSet}.\mathit{report}(#1,#2)}
\newcommand{\OptConv}{\Opt{\StaSet}.\mathit{is\_converged}()}
\newcommand{\OptReset}{\Opt{\StaSet}.\mathit{reset}()}
\newcommand{\gap}{\delta}
\newcommand{\threshold}{\gamma_{\mathrm{th}}}
\newcommand{\MaxChangeAccept}{M}

\newcommand{\IdA}{A\xspace}
\newcommand{\IdB}{B\xspace}
\newcommand{\TXa}{TX~{\IdA}\xspace}
\newcommand{\TXb}{TX~{\IdB}\xspace}
\newcommand{\TX}{TX\xspace}
\newcommand{\TXpl}{TXs\xspace}
\newcommand{\RX}{RX\xspace}
\newcommand{\stateA}{S_{\mathrm{\IdA}}}
\newcommand{\stateB}{S_{\mathrm{\IdB}}}
\newcommand{\stateAB}{S_{\mathrm{\IdA},\mathrm{\IdB}}}

\newacronym{AI}{AI}{artificial intelligence}
\newacronym{SNR}{SNR}{signal-to-noise ratio}
\newacronym{AP}{AP}{access point}
\newacronym{MIMO}{MIMO}{multiple-input multiple-output}
\newacronym{WLAN}{WLAN}{wireless local area network}
\newacronym{RSSI}{RSSI}{received signal strength indicator}
\newacronym{CSI}{CSI}{channel state information}
\newacronym{UCB}{UCB}{upper confidence bound}
\newacronym{UAV}{UAV}{unmanned aerial vehicle}
\newacronym{OFDM}{OFDM}{orthogonal frequency-division multiplexing}
\newacronym{STA}{STA}{station}
\newacronym{IoT}{IoT}{Internet of Things}
\newacronym{WebUI}{WebUI}{web user interface}
\newacronym{SDR}{SDR}{software-defined radio}
\newacronym{LOS}{LOS}{line of sight}
\newacronym{ISAC}{ISAC}{integrated sensing and communications}
\newacronym{IQR}{IQR}{interquartile range}

\begin{document}

\title{A Mechanical Antenna for Improving Capacity Fairness in Dynamic Multi-Station Scenarios\\
\thanks{This work was supported by JSPS KAKENHI Grant Number JP23K26109, JP24K20759 and the Telecommunications Advancement Foundation.
Additionally, Gemini and ChatGPT were utilized for English proofreading and stylistic refinement of this manuscript.}}

\author{\IEEEauthorblockN{1\textsuperscript{st} Akihito Taya}
\IEEEauthorblockA{\textit{Institute of Industrial Science,} \\
\textit{The University of Tokyo}\\
Tokyo, JAPAN \\
taya-a@iis.u-tokyo.ac.jp
}
\and
\IEEEauthorblockN{2\textsuperscript{nd} Yuuki Nishiyama}
\IEEEauthorblockA{\textit{Center for Spatial Information Science,} \\
\textit{The University of Tokyo}\\
Chiba, JAPAN \\
nishiyama@csis.u-tokyo.ac.jp
}
\and
\IEEEauthorblockN{3\textsuperscript{rd} Kaoru Sezaki}
\IEEEauthorblockA{\textit{Center for Spatial Information Science,} \\
\textit{The University of Tokyo}\\
Chiba, JAPAN \\
sezaki@iis.u-tokyo.ac.jp}
}

\maketitle

\begin{abstract}
While indoor \gls{IoT} and sensor networks increasingly rely on Wi-Fi \glspl{AP} to collect high-bandwidth data streams from multiple devices, conventional \glspl{AP} rely on static antenna deployments, whose fixed orientations are often suboptimal in dynamic propagation environments.
To overcome this limitation, this paper proposes a mechanical Wi-Fi antenna control system that adaptively optimizes its 3D antenna orientation for dynamic multi-station scenarios.
The proposed system autonomously actuates its physical antennas in response to perceived radio environments by combining state-specific black-box optimizers and capacity-based environment change detection.
The evaluation results show that the proposed system improves channel capacity under dynamic station combinations, avoids unnecessary re-optimization under transient blockages, and triggers re-optimization after sustained environmental changes such as continuous blockage and device relocation.
\end{abstract}
\glsresetall


\section{Introduction}
Indoor \gls{IoT} and sensor networks increasingly rely on Wi-Fi \glspl{AP} to collect high-bandwidth uplink data from cameras, LiDARs, robots, and other sensing devices.
In such networks, performance is affected not only by protocols but also by the physical radio environment, including antenna orientation, polarization mismatch, and temporary blockages.
However, conventional \glspl{AP} are typically deployed with static antenna orientations and cannot adapt to changes in active devices or propagation conditions.

While recent advances in movable and rotatable antennas \cite{zhu2024modeling,zhu2024movable,zheng2025rotatable,zheng2026rotatable} explore physical repositioning to overcome such environmental limitations, they primarily focus on fine-tuning phased array elements for narrow beams.
Similarly, although mechanical 3D beamforming systems like NetBeam \cite{bocanegra2019netbeam} exist, they rely on tightly synchronized, fully managed architectures.
Consequently, the dynamic management of broad-beam propagation---such as from omnidirectional Wi-Fi antennas in unmanaged indoor settings---remains largely unexplored, leaving these systems vulnerable to suboptimal communication performance.

\begin{figure}[!t]
\centering
\resizebox{!}{130pt}{
  \begin{tikzpicture}
    \node[anchor=south west,inner sep=0] (webui) at (0,0) {
      \includegraphics[width=0.45\textwidth,trim=8 180 7 64,clip]{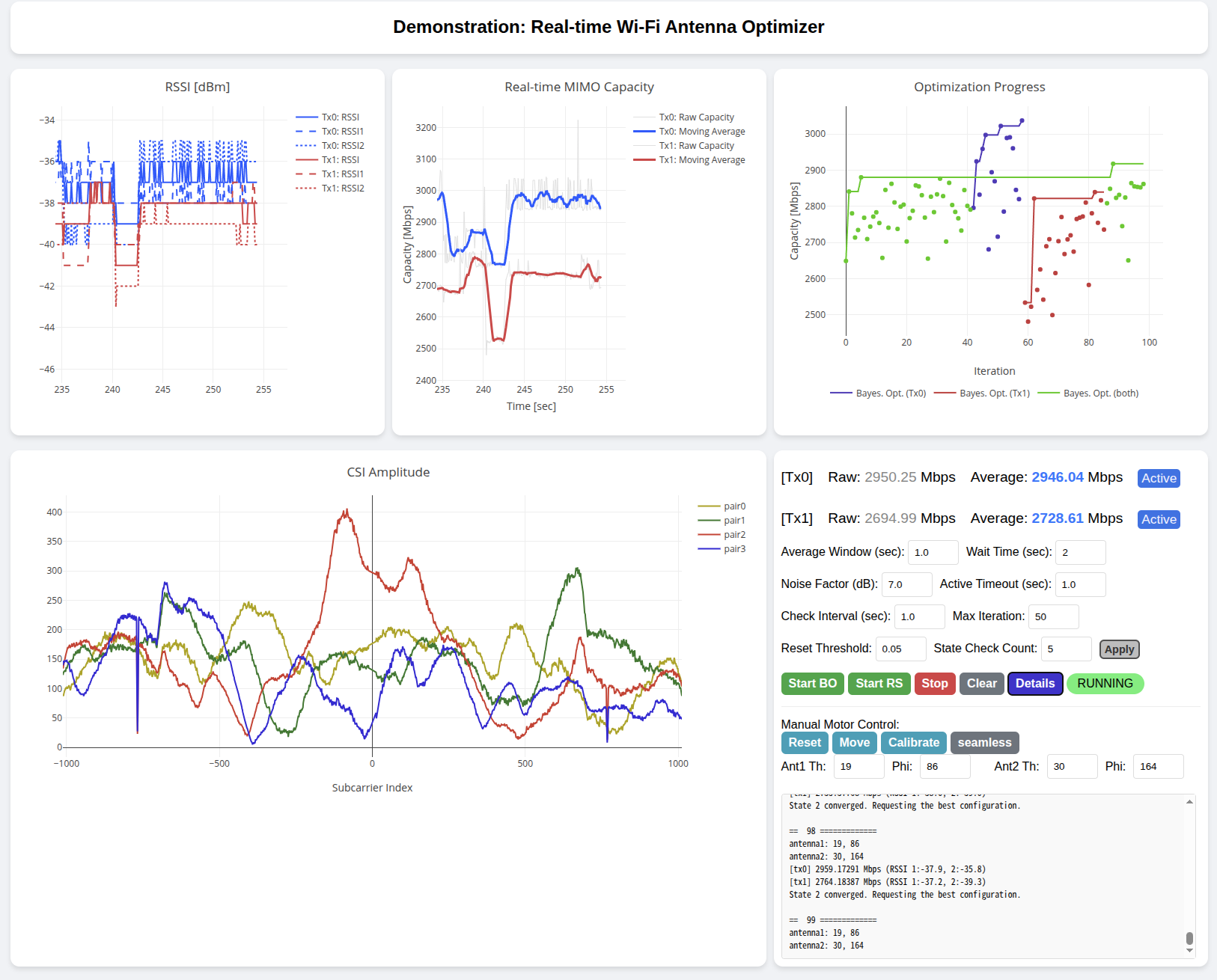}
    };
    \node[draw=black, line width=2pt, anchor=south west,inner sep=0] (device) at (6.5,0) {
      \includegraphics[width=0.15\textwidth,trim=0 0 0 0,clip]{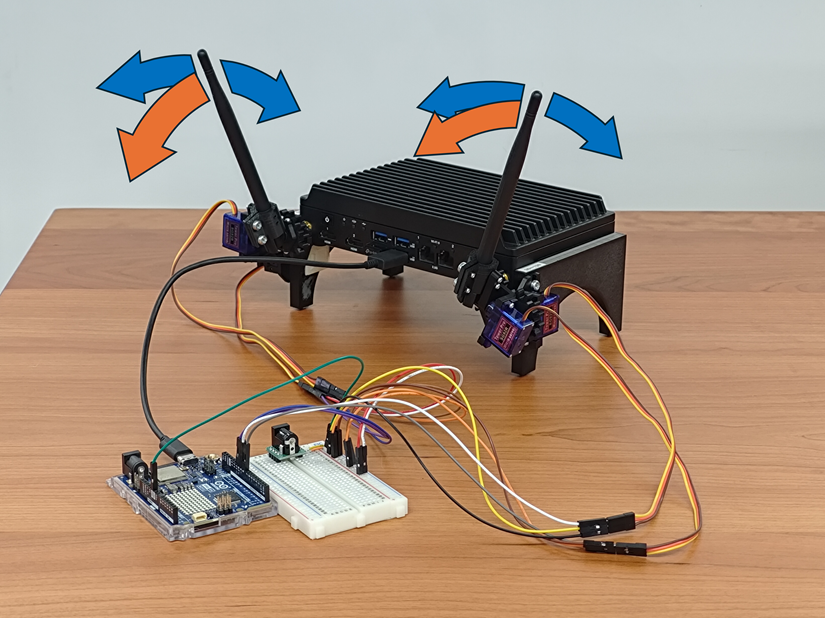}
    };
  \end{tikzpicture}
}
\caption{The proposed device equipped with mechanical Wi-Fi antennas and a web-based monitoring interface. The device can adjust the orientation of two antennas in both horizontal and vertical directions.}
\label{fig:proposed_device}
\end{figure}

To address this physical-layer bottleneck, we draw inspiration from physical \gls{AI} \cite{wu2026physical}, an emerging paradigm that integrates physics-based perception and embodied interaction to enable systems to adapt robustly via physical actuators.
Aligning with this concept, our previous work \cite{taya2026mechanical} introduced a mechanical Wi-Fi antenna (shown in Fig.~\ref{fig:proposed_device}) that automatically adjusts its 3D orientation via black-box optimization to maximize the \gls{CSI}-derived capacity for a single link.
While this preliminary system demonstrated that physical polarization matching is critical for capacity enhancement, it was limited to a single, static scenario.
Practical deployments demand a robust extension capable of managing simultaneous communications from multiple \glspl{STA} while autonomously adapting to unpredictable environmental changes.

Because the ideal antenna orientation fundamentally differs for each combination of active \glspl{STA}, the system must dynamically adapt to their activation and deactivation.
This online adaptability highlights the core advantage of mechanical antennas over traditional static deployments.
Furthermore, the system must reliably detect and react to post-optimization environmental shifts, such as device mobility or unexpected blockages.

To address these challenges, we propose a concurrent optimization framework that manages dynamic \gls{STA} combinations and reactively handles environmental shifts.
The framework maintains independent, state-specific optimizers for every possible combination of active \glspl{STA}.
It executes the optimizer corresponding to the current state while freezing the others, ensuring seamless state transitions.
Furthermore, we introduce a capacity-based environment change detection mechanism that continuously monitors the channel after convergence.
Upon detecting a sustained anomaly, it resets the optimization to adapt to the new conditions, ensuring robustness in unpredictable real-world applications.

The main contributions of this paper are summarized as follows:
\begin{itemize}
  \item We develop a mechanical Wi-Fi antenna system that autonomously tunes its 3D orientation to maximize the proportional fairness of channel capacity across multiple \glspl{STA} while adapting to environmental shifts.
  \item We design a fully asynchronous architecture to decouple processes with disparate time scales, including capacity measurement, antenna actuation, and real-time visualization. This enables the efficient, concurrent execution of optimization and environment change detection.
  \item We experimentally demonstrate the system's effectiveness in dynamic scenarios with varying active \gls{STA} combinations. The results confirm its robustness against transient \gls{LOS} blockages and adaptation to sustained changes, such as continuous blockages or device mobility.
\end{itemize}

\section{Related Work}
The traditional approach to controlling radio propagation is beamforming, which steers directivity either mechanically or electronically using phased array antennas \cite{uchendu2016survey,abbasi2019millimeter,kutty2016beamforming}.
These methods generally assume point-to-point communication with narrow beams.
In contrast, our proposed mechanical antenna controls the 3D orientation of broad-beam antennas (e.g., omnidirectional antennas) to serve multiple devices simultaneously.
Additionally, unlike conventional beamforming, which rarely addresses polarization, our device inherently manipulates polarization, a critical factor for capacity enhancement as demonstrated in \cite{taya2026mechanical}.

Recently, movable antennas \cite{zhu2024modeling,zhu2024movable} have attracted considerable attention for improving communication performance through antenna position adjustment.
These systems fine-tune the elements of a phased array to achieve more precise beamforming.
In particular, rotatable antennas \cite{zheng2025rotatable, zheng2026rotatable}, a variant of movable antennas, have also been studied for enhancing communication performance by adjusting their orientations.
However, these studies primarily focus on fine-tuning phased array elements to optimize narrow beams, effectively categorizing them under traditional beamforming.
This differs fundamentally from our approach, which independently actuates omnidirectional-like \gls{MIMO} antennas to manipulate both directivity and polarization.

In the context of multi-user communication, mechanical antenna tilting has been studied for cellular base stations to optimize coverage and capacity \cite{Niemela2004impact,athley2010impact}.
However, such large-scale antennas incur significant mechanical delay, precluding frequent reconfigurations.
Research on mechanical antenna control for indoor, small-scale \gls{WLAN} environments remains scarce.
Our system addresses this gap by employing small, rapidly actuating antennas capable of on-the-fly reconfiguration for specific active terminals.

Regarding 3D orientation control, NetBeam \cite{bocanegra2019netbeam} investigates distributed beamforming using multiple transmitters equipped with servo motors.
While providing indoor experimental insights, NetBeam relies on a fully managed architecture using \glspl{SDR} for tight synchronization and \gls{CSI} feedback, akin to cellular networks.
Our work, conversely, targets unmanaged, small-scale systems like Wi-Fi.
Crucially, our proposal introduces a comprehensive framework that asynchronously optimizes multiple active \gls{STA} combinations and reactively adapts to environmental shifts, which are not addressed by NetBeam.




\section{System Model}
We consider a scenario where a set of \glspl{STA} $\Stations$ dynamically start and stop uplink streaming sessions to a central \gls{AP}.
A typical use case is an indoor \gls{IoT} edge network in which an \gls{AP} collects high-bandwidth uplink sensor streams, such as camera, LiDAR, or robot telemetry data, from multiple sensing devices.
While we focus on the uplink, our proposed method is equally applicable to downlink scenarios by leveraging radio wave reciprocity and standard \gls{CSI} feedback mechanisms.

Unlike conventional \glspl{AP} with static antennas, our proposed \gls{AP} dynamically adjusts its 3D antenna orientation $\AngleSet$ to optimize channel capacity.
Specifically, the device independently controls two antennas in both elevation $\theta_j$ and azimuth $\varphi_j$ (for antenna $j\in\{1,2\}$), yielding a four-degree-of-freedom configuration space: $\AngleSet=(\theta_1, \varphi_1, \theta_2, \varphi_2)$.

Upon receiving data frames from \gls{STA} $i\in\Stations$ at a given orientation $\AngleSet$, the \gls{AP} measures the \gls{CSI} $\CSI_{i,\SCindex}(\AngleSet)\in\ComplexMat^{\NumRxAntenna \times \NumTxAntenna}$, where $\SCindex$, $\NumRxAntenna$ and $\NumTxAntenna$ denote the subcarrier index, the number of receive antennas, and the number of transmit antennas, respectively.
Using this \gls{CSI}, the \gls{AP} computes the achievable channel capacity, which serves as a physical-layer performance metric, and optimizes $\AngleSet$ to maximize proportional fairness among all active \glspl{STA}.

The \gls{MIMO}-\gls{OFDM} channel capacity $\Capacity_i(\AngleSet)$ between \gls{STA} $i\in\Stations$ and the \gls{AP} is calculated as \cite{goldsmith2003capacity}:
\begin{align}
  \Capacity_i (\AngleSet)
    = \BWperSubCarrier \sum_{\SCindex=1}^{\NumSubcarrier} \log_2
      \[ 
        \det \(\IdentityMat + \frac{1}{\NoiseDensity \BWperSubCarrier} \CSI_{i,\SCindex}(\AngleSet) \CSI_{i,\SCindex}(\AngleSet)^\Hermit\) 
      \], \label{eq:capacity}
\end{align}
where $\BWperSubCarrier$, $\NumSubcarrier$, and $\NoiseDensity$ denote the subcarrier bandwidth, the number of subcarriers, and the noise density per antenna, respectively.
Additionally, $\IdentityMat$ is the $\NumRxAntenna \times \NumRxAntenna$ identity matrix and $[\cdot]^\Hermit$ denotes the Hermitian transpose.
Note that $\CSI_{i,\SCindex}(\AngleSet)$ is treated as an effective channel matrix that includes the received signal power scaling.

To balance overall channel capacity and promote fairness among users, we adopt proportional fairness \cite{shi2014fairness} based on the time-averaged capacity $\AvgCapacity_i(\AngleSet)$ over a window $\AvgTime$ as our optimization objective.
We define the proportional fairness metric $\Fairness_{\StaSet}(\AngleSet)$ for the active stations $\StaSet\subseteq\Stations$ as:
\begin{align}
  \Fairness_{\StaSet}(\AngleSet) \defeq \sum_{i \in \StaSet} \log \(\AvgCapacity_i (\AngleSet)\). \label{eq:fairness}
\end{align}
Since the ideal antenna orientation inherently depends on which devices are active, the system aims to find the optimal configuration $\AngleSetIdeal_\StaSet \defeq \argmax_{\AngleSet} \Fairness_{\StaSet}(\AngleSet)$ for each unique \gls{STA} combination $\StaSet$.

Practically, commercial hardware often reports a normalized \gls{CSI}, $\NormCSI_{i,\SCindex}(\AngleSet) = \gainCoef \CSI_{i,\SCindex}(\AngleSet)$.
Although the scaling factor $\gainCoef$ is typically undisclosed, it can be derived from the \gls{RSSI}.
Because \gls{RSSI} encapsulates the total signal and noise power across all subcarriers, it satisfies:
\begin{align}
  \rssi = \sum_{\SCindex=1}^{\NumSubcarrier} \|\CSI_{i,\SCindex}(\AngleSet)\|^2 + \NumSubcarrier \NumRxAntenna \NoiseDensity \BWperSubCarrier,
\end{align}
where $\|\cdot\|$ denotes the Frobenius norm.
Consequently, $\gainCoef$ is extracted as:
\begin{align}
  \gainCoef^2 = \frac{\sum_{\SCindex=1}^{\NumSubcarrier} \|\NormCSI_{i,\SCindex}(\AngleSet)\|^2}{\rssi - \NumSubcarrier \NumRxAntenna \NoiseDensity \BWperSubCarrier}.
\end{align}
Substituting this into \eqref{eq:capacity} yields the practical capacity formula:
\begin{align}
  \Capacity_i (\AngleSet)
    &= \BWperSubCarrier \nonumber \\
    &\times \sum_{\SCindex=1}^{\NumSubcarrier} \log_2
      \[ 
        \det \(\IdentityMat + \frac{1}{\gainCoef^2 \NoiseDensity \BWperSubCarrier} \NormCSI_{i,\SCindex}(\AngleSet) \NormCSI_{i,\SCindex}(\AngleSet)^\Hermit\) 
      \]. \label{eq:capacity-measurement}
\end{align}

\section{Concurrent Optimization Framework and Environment Change Detection}
\subsection{Asynchronous System Architecture} \label{sec:async_architecture}
\begin{figure}[!t]
  \centering
  \resizebox{!}{180pt}{
\newcommand{\getdata}[6]{
  \draw[->, thick] (#1, #3) -- node[above, font=\footnotesize, align=center] {#5} (#2, #3);
  \draw[<-, thick, dashed] (#1, #3-#4) -- node[below, font=\footnotesize, align=center] {#6} (#2, #3-#4);
}
\newcommand{\senddata}[4]{
  \draw[->, thick] (#1, #3) -- node[above, font=\footnotesize, align=center] {#4} (#2, #3);
}
\begin{tikzpicture}[>=stealth, font=\sffamily\small]
  \tikzset{
    module/.style={
      draw,
      fill=yellow!20,
      rounded corners=2pt,
      minimum width=1.0cm,
      minimum height=0.6cm,
      align=center,
    },
    storage/.style={
      cylinder,
      draw,
      fill=yellow!20,
      minimum height=0.8cm,   
      minimum width=1.5cm,    
      cylinder uses custom fill,
      aspect=0.2, 
      rotate=90, 
      align=center,
    },
    sleep/.style={
      draw,
      fill=white,
      minimum width=0.2cm,
      minimum height=#1,
      inner sep=0pt,
      anchor=north,
    }
  }

  \def\txx{0.1}
  \def\csix{1.7}
  \def\bufferx{3.8}
  \def\managerx{5.9}
  \def\optx{7.45}
  \def\actx{8.5}

  \def\mody{0.0}
  \def\suby{-1.2}

  \node[module]  (TX)      at (\txx, \mody) {STA};
  \node[module]  (CsiMod)  at (\csix, \mody) {Capacity\\Calculator};
  \node[storage] (Buffer)  at (\bufferx, \suby) {\rotatebox{-90}{Buffer}}; 
  \node[module]  (Manager) at (\managerx, \mody) {Optimization\\Manager};
  \node[module]  (OptMod)  at (\optx, \suby) {Optimizer};
  \node[module]  (ActMod)  at (\actx, \mody) {Actuator};

  \def\braceYshift{0.2}
  \draw[decorate, decoration={brace, amplitude=6pt}, thick]
    ([yshift=\braceYshift cm]CsiMod.north west) -- ([yshift=\braceYshift cm]ActMod.north east)
    node[midway, above=6pt] {AP with Mechanical Antenna};

  \def\vlineS{-0.3}
  \def\vlineSdouble{-0.4}
  \def\vlineSsub{-1.5}
  \def\vlineE{-7.3}
  \draw[thick] (\txx, \vlineS) -- (\txx, \vlineE);
  \draw[thick] (\csix, \vlineSdouble) -- (\csix, \vlineE);
  \draw[thick] (\bufferx, \vlineSsub) -- (\bufferx, \vlineE);
  \draw[thick] (\managerx, \vlineSdouble) -- (\managerx, \vlineE);
  \draw[thick] (\optx, \vlineSsub) -- (\optx, \vlineE);
  \draw[thick] (\actx, \vlineS) -- (\actx, \vlineE);

  \def\ystart{-0.6}
  \def\ydiff{1.5}
  \def\measureDelay{0.2}
  \foreach \i in {1,...,4} {
    \def\y{\ystart-\i*\ydiff}
    \def\delayedY{\y-\measureDelay}
    \senddata{\txx}{\csix}{\y}{Data}
    \senddata{\csix}{\bufferx}{\delayedY}{Capacity/RSSI}
  }
  
  \def\loopLeftX{\bufferx-0.5}
  \def\loopRightX{\actx+0.5}
  \def\loopTopY{-2.6}
  \def\loopBottomY{-6.9}
  \draw[dotted, blue, thick] 
    (\loopLeftX, \loopTopY) rectangle (\loopRightX, \loopBottomY);
  \node[blue, draw=blue!60, fill=blue!5, anchor=north west] at (\loopLeftX, \loopTopY) {optimization loop};

  \def\requestDelay{0.1}
  \def\activeCheckY{-4.0}
  \getdata{\managerx}{\bufferx}{\activeCheckY}{\requestDelay}{Check}{active STA}

  \def\suggestionY{\activeCheckY-0.3}
  \def\setAngleY{\suggestionY-1.0}
  \getdata{\managerx}{\optx}{\suggestionY}{\requestDelay}{Request}{suggestion}
  \senddata{\managerx}{\actx}{\setAngleY}{Set angle $\AngleSet$}

  \def\waitStartY{\setAngleY}
  \def\waitDuration{0.8}
  \def\waitEndY{\waitStartY-\waitDuration}
  \node[sleep=\waitDuration cm] (Wait) at (\managerx, \waitStartY) {};
  \node[right, font=\footnotesize, align=left] at ([yshift=2]Wait.east) {Wait $\WaitTime$};

  \def\readCsiY{\waitEndY}
  \def\reportY{\readCsiY-0.3}
  \getdata{\managerx}{\bufferx}{\readCsiY}{\requestDelay}{Request}{average capacity}
  \senddata{\managerx}{\optx}{\reportY}{Report}

\end{tikzpicture}}
  \caption{Sequence diagram of the asynchronous antenna tuning framework. The high-frequency capacity measurement is decoupled from the slow optimization loop to enable online adaptation.}
  \label{fig:sequence_diagram}
\end{figure}
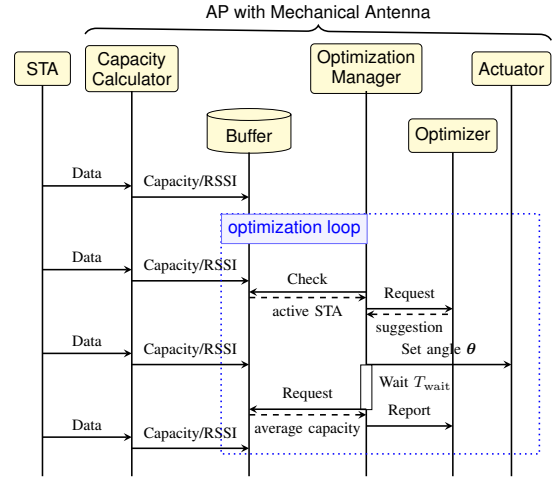

In realistic unmanaged networks like Wi-Fi, the \gls{AP} cannot dictate the exact transmission timings of the active \glspl{STA}.
If the system were to perform high-frequency capacity computation and mechanical antenna actuation, which requires hundreds of milliseconds, sequentially within a synchronous loop, the overall processing efficiency would severely degrade.
To resolve this bottleneck, we designed a fully asynchronous architecture, as illustrated in Fig.~\ref{fig:sequence_diagram}.
The system decouples the fast sensing processes from the relatively slow optimization control loop.
A background Capacity Calculator continuously processes incoming frames and updates the moving average of the channel capacity in a shared memory buffer.
Independently, the Optimization Manager queries this buffer only when evaluating the objective function.
This thread-safe, low-latency data exchange completely eliminates the need to synchronize packet arrivals with mechanical actuation, ensuring efficient adaptability and facilitating smooth system monitoring via real-time visualization.

\subsection{State-Specific Concurrent Optimization} \label{sec:state_specific_opt}
Because the joint optimization of multiple 3D antenna orientations to maximize proportional fairness among multiple \glspl{STA} can be highly complex and potentially non-convex, we formulate the tuning process as a black-box optimization problem.
While our proposed architecture is solver-agnostic and can integrate any black-box optimization algorithm, methods that require a large number of agents to explore the parameter space are less suitable for our system due to the sequential nature of physical measurements.
Consequently, we employ Bayesian optimization \cite{garnett2023bayesian} in our evaluations because it is well suited for optimizing expensive-to-evaluate functions with a limited number of samples.

The primary challenge in dynamic scenarios is that the optimal antenna configuration fundamentally shifts whenever a \gls{STA} becomes active or inactive.
To seamlessly manage these transitions, the Optimization Manager maintains a set of independent, state-specific optimizers, denoted as $\mathcal{O}[\StaSet]$, for every possible active \gls{STA} combination $S$.

\begin{figure}[!t]
  \begin{algorithm}[H]
    \caption{Concurrent antenna orientation optimization for multiple \gls{STA} combinations} \label{alg:proposed_framework}
    \begin{algorithmic}[1]
      \While{optimization is running}
        \State Observe the current active \gls{STA} combination $\StaSet$
        \OnelineIf{$\StaSet = \emptyset$} Wait $\WaitTime$ and \textbf{continue}
        \State $\AngleSet \gets \OptSuggest$
        \State Move antennas to $\AngleSet$ and wait $\WaitTime$
        \OnelineIf{any \gls{STA} in $\StaSet$ is inactive} \textbf{continue} \label{algln:verify_active}
        \ForAll {$i \in \StaSet$}
          \State Calculate average channel capacity $\AvgCapacity_i(\AngleSet)$
        \EndFor
        
        \If {not $\OptConv$}
          \State Calculate the objective value $\Fairness_{\StaSet}(\AngleSet)$ by \eqref{eq:fairness}
          \State $\OptReport{\AngleSet}{\Fairness_{\StaSet}(\AngleSet)}$
          \If {converged at this round}
            \State Store the best antenna orientation $\AngleSetBest_\StaSet$
            \State $\CapBest_{\StaSet,i} \gets \AvgCapacity_i(\AngleSetBest_\StaSet)\ (\forall i \in \StaSet)$
          \EndIf
        \Else
          \If {environment change is detected}
            \State $\OptReset$
          \EndIf
        \EndIf
      \EndWhile
    \end{algorithmic}
  \end{algorithm}
\end{figure}

Algorithm~\ref{alg:proposed_framework} details the execution flow of the proposed concurrent optimization framework.
At the beginning of each iteration, the Optimization Manager observes the active \gls{STA} combination $\StaSet$.
If $\StaSet$ is not empty, the manager invokes only the corresponding optimizer $\Opt{\StaSet}$.
Crucially, all other optimizers are frozen, preserving their learned internal states until their specific \gls{STA} combinations become active again.

The selected optimizer then suggests an antenna configuration $\AngleSet$.
Because physically adjusting the mechanical antennas takes on the order of hundreds of milliseconds, the manager commands the Actuator to move the antennas to $\AngleSet$ and waits for a stabilization period $\WaitTime$ before measuring the updated channel capacity.
During this waiting phase, the active \gls{STA} combination might dynamically change (e.g., a station might finish its transmission).
To handle this dynamic event, the manager explicitly verifies whether all \glspl{STA} in $\StaSet$ are still active after the delay (Line~\ref{algln:verify_active}).
If any \gls{STA} is found to be inactive, the current optimization round is safely aborted to avoid injecting mismatched capacity evaluations into the optimizer.

Otherwise, the manager retrieves the average capacities $\AvgCapacity_i(\AngleSet)$ from the shared buffer, calculates the proportional fairness objective, and reports it to update $\Opt{\StaSet}$.
Upon convergence, the manager stores the optimal orientation $\AngleSetBest_\StaSet$ and the corresponding achieved capacities $\CapBest_{\StaSet,i}$, which serve as the baseline for the environment change detection described in the next subsection.

The proposed framework maintains independent optimizers for each possible combination of active \glspl{STA}.
Thus, the number of possible non-empty states grows as $2^{|\Stations|} - 1$.
Consequently, the storage required to maintain the learned optimizer states and the cumulative time required to optimize all possible states may become significant as the number of \glspl{STA} increases.
The current framework therefore targets small-scale deployments, such as homes and small offices.
Extending the framework to handle larger numbers of \glspl{STA} is left for future work.

\subsection{Capacity-Based Environment Change Detection}
After convergence, the optimizer $\Opt{\StaSet}$ is designed to consistently output the optimal orientation $\AngleSetBest_\StaSet$, allowing the Optimization Manager to seamlessly continue its evaluation loop.
However, the manager must continuously monitor the links to detect post-optimization environmental shifts, such as device mobility or physical blockages.

Relying directly on raw \gls{CSI} for anomaly detection is impractical; \gls{CSI} matrices are high-dimensional complex matrices whose phases often exhibit unstable fluctuations even in strictly static environments.
Instead, we adopt the channel capacity as a robust, macroscopic feature to reliably characterize the radio environment.
The Optimization Manager calculates the relative capacity deviation $\gap_i$ between the currently observed average capacity $\AvgCapacity_i(\AngleSet)$ and the recorded baseline $\CapBest_{\StaSet,i}$, defined as $\gap_i \defeq |\AvgCapacity_i(\AngleSet) - \CapBest_{\StaSet,i}| / \CapBest_{\StaSet,i}$.

To prevent the system from overreacting to transient blockages (e.g., a person simply walking past the \gls{AP}), the manager increments a state-specific counter if $\gap_i$ for any active \gls{STA} exceeds a predefined threshold $\threshold$.
If this condition is met for $\MaxChangeAccept$ consecutive iterations, the manager concludes that a sustained environmental change has occurred and resets $\Opt{\StaSet}$ to dynamically adapt to the new conditions.

\section{Evaluations} \label{sec:evaluation}

\subsection{Static Scenario Evaluation}
\newcommand{\statsSize}{0.43\textwidth}
\begin{figure}[!t]
  \centering
  \begin{tikzpicture}
    \node[anchor=north,inner sep=0] (loc) at (0, 0) {
      \subfloat[Experimental setup.]{
        \resizebox{!}{90pt}{\begin{tikzpicture}
  \tikzset{
    arrow/.style={
      thin,
      dashed,
      {Stealth[length=7,width=6]}-{Stealth[length=7,width=6]}
    },
    eventLabel/.style={
      font=\footnotesize\sffamily,
      color=#1,
      align=left
    },
    device/.style={
      draw,
      thick,
      minimum width=30pt,
      minimum height=13pt
    }
  }
  \def\left{0.4}
  \def\right{7.0}
  \def\top{3.6}
  \def\bottom{-0.1}

  \node[inner sep=0] (BL) at (\left, \bottom) {};
  \node[inner sep=0] (TL) at (\left, \top) {};
  \node[inner sep=0] (BR) at (\right, \bottom) {};
  \node[inner sep=0] (TR) at (\right, \top) {};

  \node[device] (RX) at (1.04, 3.45) {\RX};
  \draw[line width=7pt] (0.5, 1.9) -- (0.6, 2.8);
  \draw[line width=7pt] (1.6, 1.9) -- (1.5, 2.8);
  \draw[line width=7pt] (0.5, 2.8) -- (1.6, 2.8);
  \draw[line width=7pt] (0.5, 2.4) -- (1.6, 2.4);

  \node[anchor=south west,align=left] at (0.3, 3.6) {
    RX and TX are mounted on a ladder at a height of 1.7\,m\\and on desks at a height of 0.7\,m, respectively.
   };

  \draw[thin] (1.8, 1.8) rectangle (3.0, 2.9);
  \draw[thin] (5.2, 0.0) rectangle (6.9, 2.9);
  \node[inner sep=2] at (6.9, 0.0) [anchor=south east] {Desk};

  \def\RxCenter{1.05};
  \def\TxACenter{5.8};
  \def\TxAHeight{2.5};
  \def\TxBCenter{2.4};
  \def\TxBNextHeight{0.9};
  \def\LineBottomA{0.6}
  \def\LineBottomB{0.6}
  \def\ArrowAheight{1.0};
  \def\ArrowBheight{1.4};

  \node[device] (TXa) at (\TxACenter, \TxAHeight) {\TXa};
  \node[device] (TXb) at (\TxBCenter, \TxAHeight) {\TXb};
  \node[device] (TXbNext) at (\TxACenter, \TxBNextHeight) {\TXb};
  \node[inner sep=2] at (TXb.east) [anchor=west] {Layout 1};
  \node[inner sep=2] at (TXbNext.east) [anchor=west] {Layout 2};

  \draw[draw, thin] (RX.south) -- (\RxCenter, \LineBottomA);
  \draw[draw, thin] (TXb.south) -- (\TxBCenter, \LineBottomB);
  \draw[arrow] (\RxCenter, \ArrowAheight) -- (\TxBCenter, \ArrowAheight) node[midway, below, inner sep=2] {0.7\,m};
  \draw[arrow] (\RxCenter, \ArrowBheight) -- (\TxACenter, \ArrowBheight) node[midway, below, inner sep=2] {2.2\,m};
  \draw[arrow] (TXa.south) -- (TXbNext.north) node[midway, right] {1.3\,m};
\end{tikzpicture}
        \label{fig:boxplot_setup}}
      }
    };
    \node[anchor=north,inner sep=0] (fig1) at (loc.south) {
      \subfloat[Layout 1]{
        \includegraphics[width=\statsSize,trim=5 5 5 5,clip]{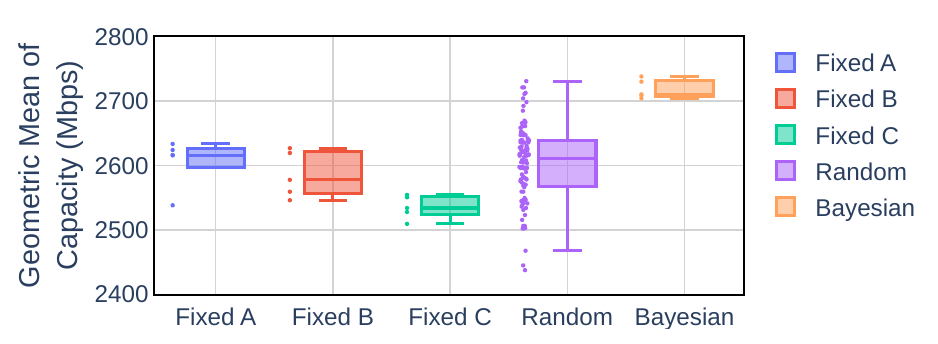}
        \label{fig:boxplot_layout1}
      }
    };
    \node[anchor=north,inner sep=0] (fig2) at (fig1.south) {
      \subfloat[Layout 2]{
        \includegraphics[width=\statsSize,trim=5 5 5 5,clip]{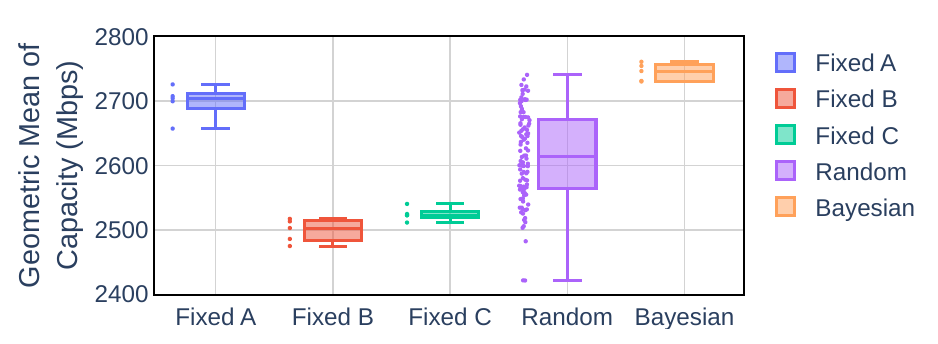}
        \label{fig:boxplot_layout2}
      }
    };
  \end{tikzpicture}
  \caption{
    Distribution of the geometric mean of the channel capacities in static scenarios. (a) Experimental setup showing the locations of the TXs and RX. (b) and (c) compare fixed, randomly selected, and Bayesian-optimized antenna orientations under two layouts with different \TXa antenna orientations and different \TXb locations. Bayesian optimization consistently finds high-performing orientations in both layouts.
  }
  \label{fig:boxplot}
\end{figure}

We first evaluated the proposed system in static scenarios to assess the effectiveness of the optimization framework.
The evaluation was conducted in an indoor environment under two layouts differing in \TXa antenna orientations and \TXb locations, as shown in Fig.~\ref{fig:boxplot_setup}.
For each layout, we compared three representative fixed antenna configurations, randomly selected orientations, and Bayesian-optimized orientations.
The four-dimensional antenna orientations $(\theta_1,\varphi_1,\theta_2,\varphi_2)$ of Fixed A, B, and C were set to $(0^\circ,90^\circ,0^\circ,90^\circ)$, $(90^\circ,90^\circ,90^\circ,90^\circ)$, and $(90^\circ,180^\circ,90^\circ,180^\circ)$, respectively, where $\theta$ represents the elevation angle ranging from $0^\circ$ (vertical) to $90^\circ$ (horizontal), and $\varphi$ represents the azimuth angle ranging from $60^\circ$ (toward the center of the \RX) to $180^\circ$ (outward from the \RX).
Each fixed orientation configuration was evaluated five times.
Random orientations were uniformly and independently sampled 100 times from the feasible orientation space to simulate arbitrary antenna configurations selected by non-expert users.
Bayesian optimization was independently repeated five times, with 50 optimization iterations in each trial.
For each trial, the best orientation found within the 50 iterations was used for comparison.
As the performance metric, we used the geometric mean of the channel capacities of the two links, whose maximization is equivalent to maximizing the proportional-fairness objective in \eqref{eq:fairness}.
The resulting distributions are shown in Figs.~\ref{fig:boxplot_layout1} and~\ref{fig:boxplot_layout2}.
The boxes represent the median and \gls{IQR}, with whiskers defined using Tukey's $1.5\times$\gls{IQR} rule.
Random orientations can occasionally achieve performance comparable to Bayesian optimization, whereas Bayesian optimization consistently identifies high-performing orientations.
The results also show that Bayesian optimization found high-performing orientations within 50 iterations in both static layouts.

\subsection{Experimental Setup of Dynamic Scenarios}
\begin{figure}[!t]
  \centering
  \resizebox{!}{110pt}{\begin{tikzpicture}
  \tikzset{
    arrow/.style={
      thin,
      dashed,
      {Stealth[length=7,width=6]}-{Stealth[length=7,width=6]}
    },
    eventLabel/.style={
      font=\footnotesize\sffamily,
      color=#1,
      align=left
    }
  }
  \def\left{-0.2}
  \def\right{10.5}
  \def\top{4.7}
  \def\bottom{0}

  \node[inner sep=0] (BL) at (\left, \bottom) {};
  \node[inner sep=0] (TL) at (\left, \top) {};
  \node[inner sep=0] (BR) at (\right, \bottom) {};
  \node[inner sep=0] (TR) at (\right, \top) {};

  \node[draw, thick, minimum width=30pt, minimum height=13pt] (RX) at (1.04, 3.1) {\RX};
  \draw[line width=7pt] (0.5, 1.9) -- (0.6, 2.8);
  \draw[line width=7pt] (1.6, 1.9) -- (1.5, 2.8);
  \draw[line width=7pt] (0.5, 2.8) -- (1.6, 2.8);
  \draw[line width=7pt] (0.5, 2.4) -- (1.6, 2.4);

  \node[anchor=north west,align=left] at (0.3, 4.6) {RX is mounted on a ladder\\at a height of 1.7\,m.};
  \node[anchor=north east,align=left] at (10.3, 3.4) {TXs are placed on\\desks at a height\\of 0.7\,m.};

  \draw[thin] (0.4, 0.2) rectangle (4.8, 1.8);
  \draw[thin] (6.2, 0.2) rectangle (9.9, 1.8);
  \node[] at (9.9, 0.2) [anchor=south east] {Desk};

  \node[draw, thick, minimum width=30pt, minimum height=13pt] (TXa) at (6.8, 1.5) {\TXa};
  \node[draw, thick, minimum width=30pt, minimum height=13pt] (TXb) at (1.04, 1.5) {\TXb};
  \node[draw, fill=black!20, thick, minimum width=30pt, minimum height=13pt] (TXbNext) at (4.2, 0.8) {\TXb};
  
  \draw[draw=red, -{Latex[length=7,width=10]}, line width=3pt] (TXb) to[bend left=7] node[above, eventLabel=red, xshift=23pt, yshift=-34pt] {Moved at $t=790$\,s} (TXbNext);

  \node (man) at (5.75, 0.4) {};
  \draw[draw, thin] (man) circle (0.38 and 0.25);
  \draw[draw, thin] (man) circle (0.20 and 0.20);
  \draw[draw=blue, -{Latex[length=7,width=6]}, line width=2pt] (5.9, 0.8) -- (5.9, 3.5) to[bend right=60] (5.6, 3.5) -- (5.6, 0.8);
  \node[eventLabel=blue, anchor=south west] at ([xshift=-20pt]5.75, 3.7) {A subject walks across the LOS\\between \TXa and RX ($t=460$--$510$\,s).};

  \draw[draw, thin] (TXa.north) -- (6.8, 3.5);
  \draw[draw, thin] (TXb.east) -- (2.5, 1.5);
  \draw[draw, thin] (TXb.south) -- (1.04, 0.7);
  \draw[arrow] (RX.east) -- (6.8, 3.1) node[midway, above] {2.2\,m};
  \draw[arrow] (2.2, 1.5) -- (2.2, 3.1) node[midway, right] {0.7\,m};
  \draw[arrow] (TXbNext.north) -- (4.2, 3.1) node[midway, right, yshift=7pt] {1.0\,m};
  \draw[arrow] (TXbNext.west) -- (1.04, 0.8) node[midway, above] {1.3\,m};
\end{tikzpicture}}
  \caption{
    Experimental setup and spatial layout.
    The figure details the physical arrangement of the RX and TXs, alongside the specific dynamic events (a pedestrian LOS blockage and the physical relocation of TX B).
  }
  \label{fig:location}
\end{figure}
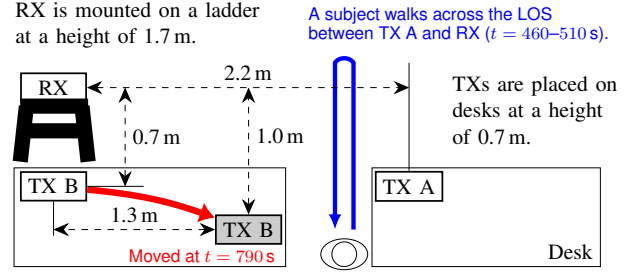

We evaluated the proposed device in an indoor environment with two \TXpl as shown in Fig.~\ref{fig:location}.
The \gls{AP} independently controlled two omnidirectional antennas in azimuth ($[60^\circ, 180^\circ]$) and elevation ($[0^\circ, 90^\circ]$).
Using an ASUS NUC 13 Rugged (\RX, \TXa) and an Intel NUC 12 Pro Kit (\TXb), we used PicoScenes \cite{jiang2022eliminating} to emulate uplink streaming.
Instead of actual application payloads, the \TXpl injected \gls{CSI} measurement frames every 0.1\,s.
The \RX flagged a \TX as inactive if no frames arrive for 2.0\,s.
By manually toggling the \TXpl, we tested three active \gls{STA} combinations: $\stateA$, $\stateB$, and $\stateAB$ (both active).

For Bayesian optimization, we set a maximum of 50 iterations per state.
To ensure mechanical stabilization before capacity evaluation, we set the wait time $\WaitTime=2$\,s and measured the average capacity over a window $\AvgTime=1$\,s.
The environment change detection parameters were configured to $\threshold=0.05$ and $\MaxChangeAccept=5$.
Finally, the noise density was calculated as $\NoiseDensity=\Boltzman \Temperature \NoiseFactor$, where $\Boltzman$, $\Temperature$, and $\NoiseFactor$ denote the Boltzmann constant, the temperature in Kelvin, and the noise figure of the receiver, respectively.
We set $\Temperature=300$\,K and $\NoiseFactor=7$\,dB, which are typical values for Wi-Fi receivers \cite{merlin2015tgax}.

\subsection{Applicability to Dynamic Scenarios}
\begin{figure*}[!t]
  \centering
  \resizebox{0.83\textwidth}{!}{\input{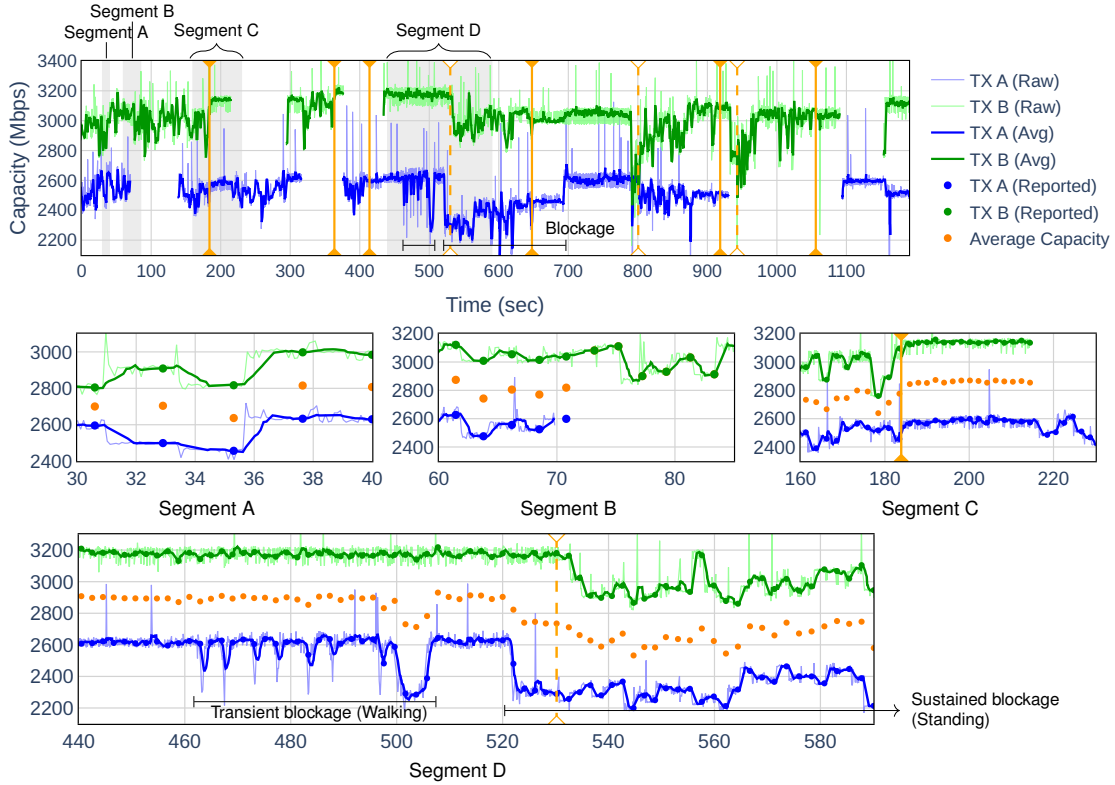}}
  \caption{
    Real-time trace of the channel capacities during the emulated streaming session.
    The light and dark solid lines represent the raw $\Capacity_i(\AngleSet)$ and moving-average $\AvgCapacity_i(\AngleSet)$ capacities, respectively.
    The blue and green colors correspond to \TXa and \TXb.
    Circle markers indicate the capacity values reported to the optimizers, while orange markers show the average capacity.
    Vertical solid and dashed orange lines denote optimization completions and resets, respectively.
    Segments A--D highlight the system's dynamic adaptation to specific events.
  }
  \label{fig:timeline}
\end{figure*}

Fig.~\ref{fig:timeline} illustrates the time series of the channel capacities for \TXa and \TXb.
Initially, both \TXpl are active, prompting the proposed device to explore antenna orientations to maximize proportional fairness using $\Opt{\stateAB}$.
Segment~A (30--40\,s) highlights the fundamental optimization loop: while the raw capacity fluctuates significantly every $\WaitTime=2$\,s due to mechanical actuations, the averaging over $\AvgTime=1$\,s provides a stable and reliable metric for the optimizer.

The system seamlessly manages dynamic \gls{STA} states.
At $t=70$\,s, \TXa becomes inactive (Segment~B); the system suspends $\Opt{\stateAB}$ and invokes $\Opt{\stateB}$.
When \TXa resumes at $140$\,s, $\Opt{\stateAB}$ restarts from its suspended state, eventually completing at $183$\,s (Segment~C).
After the completion, the capacities stabilize.
Because the optimized orientation $\AngleSetBest_{\stateAB}$ is securely cached, the system instantly restores it whenever the state returns to $\stateAB$ (e.g., at $297$\,s and $436$\,s).
Additionally, when \TXb becomes inactive at $216$\,s, the manager correctly switches to $\Opt{\stateA}$ to optimize the single link for \TXa.

After demonstrating the initial optimization, we evaluated the environment change detection algorithm by blocking the \gls{LOS}.
From $460$ to $510$\,s, a subject repeatedly walked back and forth across the \gls{LOS} of \TXa (Fig.~\ref{fig:location}).
As Segment~D shows, this caused severe capacity drops.
Crucially, because our detection algorithm requires a \emph{sustained} change over $\MaxChangeAccept$ iterations, it successfully ignored this transient blockage without prematurely resetting the optimizer.
Conversely, when the subject continuously blocked the \gls{LOS} starting at $520$\,s, the system detected the sustained degradation and properly reset $\Opt{\stateAB}$ at $530$\,s (dashed orange line).
After the optimization of $\Opt{\stateAB}$ completed at $650$\,s, the channel capacities stabilized again.
Here, although the proposed device could maintain high capacity for \TXb, the optimizer correctly prioritized improving the lower-capacity link (\TXa) to increase the proportional fairness.
Note that the subject cleared the \gls{LOS} at $690$\,s, but the system did not reset $\Opt{\stateAB}$ because the relative capacity gap did not exceed the threshold $\threshold=0.05$.

Finally, at $790$\,s, we physically relocated \TXb to a new position (gray box in Fig.~\ref{fig:location}).
This sustained environmental shift triggered an immediate reset of $\Opt{\stateAB}$ at $800$\,s.
Because only $\Opt{\stateAB}$ was updated during this period, the cached baseline for $\Opt{\stateB}$ became obsolete.
Consequently, when \TXa became inactive at $940$\,s, the system correctly detected the baseline mismatch and reset $\Opt{\stateB}$ to re-adapt \TXb to its new environment.


Overall, the proposed dynamic mechanical antenna framework exhibited substantial capacity variations across antenna configurations, with differences of approximately 200\,Mbps for \TXa and 400\,Mbps for \TXb between the worst and best configurations.
As shown in the static scenario evaluation, Bayesian optimization found high-performing orientations within 50 iterations, requiring only a few minutes for each state.
Although this timescale is not fast enough to track continuous station mobility or frequent environmental changes, the proposed system is suitable for quasi-static indoor scenarios, such as desk work, video conferencing, and video streaming.
The environmental change detection mechanism further enhances the practical feasibility of the system by avoiding unnecessary re-optimization in response to transient events.
Furthermore, proportional fairness balances the channel capacities among active \glspl{STA}, demonstrating the effectiveness of the proposed framework in dynamic multi-station indoor environments.

\subsection{Impact of Antenna Orientation on RSSI}
\newcommand{\contourSize}{0.23\textwidth}
\begin{figure}[!t]
  \centering
  \begin{tikzpicture}
    \node[anchor=south west,inner sep=0] (fig) at (0,0) {
      \begin{tabular}{@{}c@{}c@{}}
        \subfloat[Initial environment]{
          \includegraphics[width=\contourSize,trim=10 10 10 10,clip]{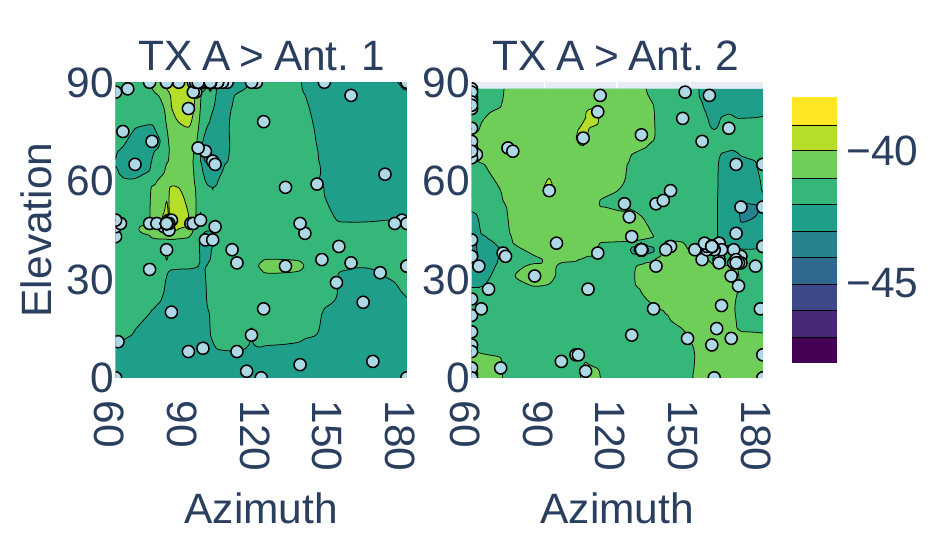}
          \includegraphics[width=\contourSize,trim=10 10 10 10,clip]{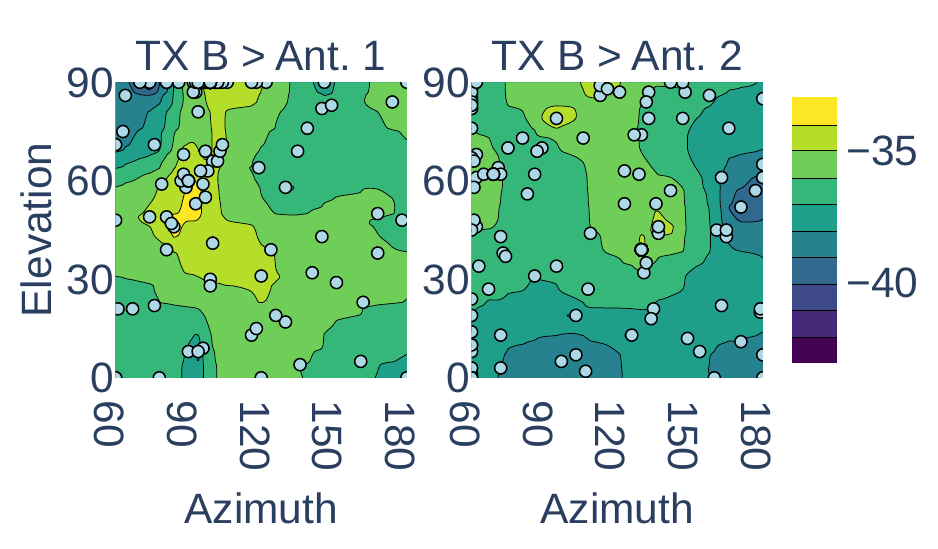}
        } \\
        \subfloat[During \gls{LOS} blockage of \TXa]{
          \includegraphics[width=\contourSize,trim=10 10 10 10,clip]{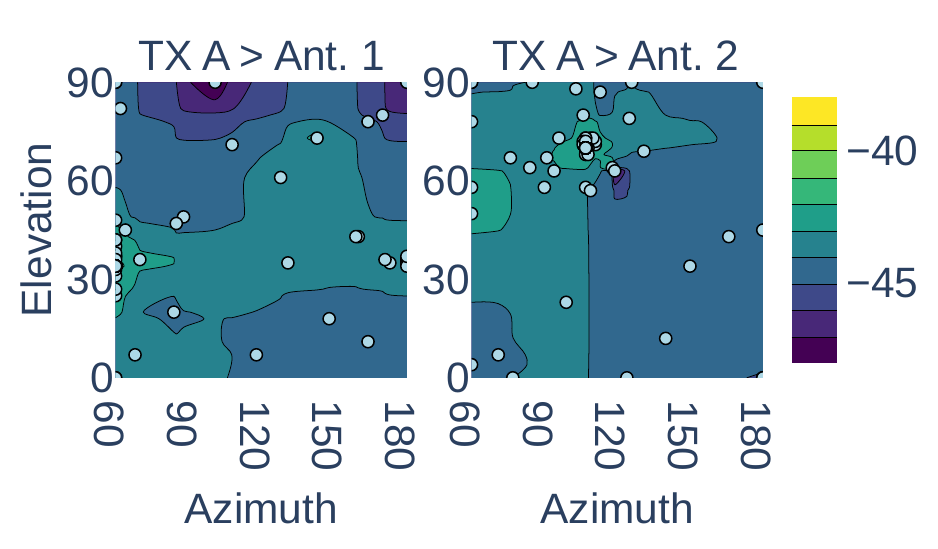}
          \includegraphics[width=\contourSize,trim=10 10 10 10,clip]{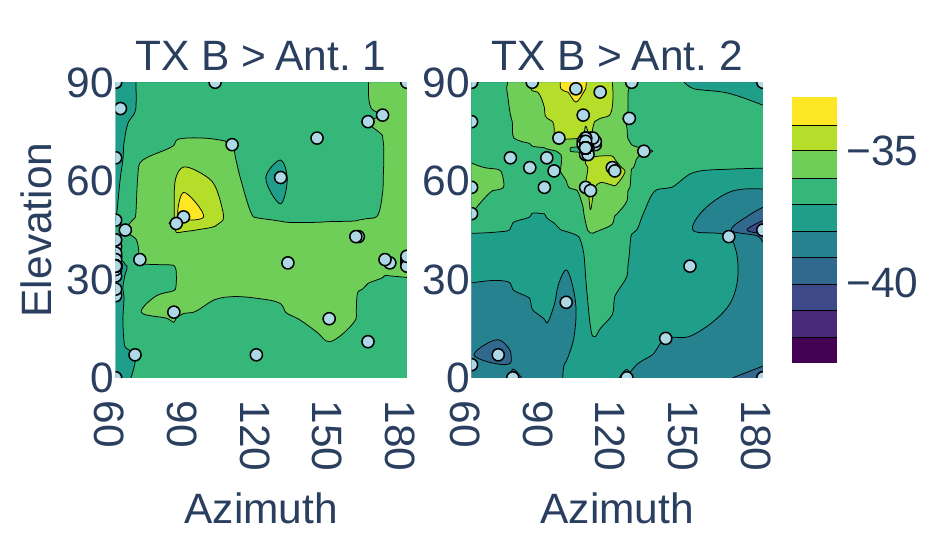}
        } \\
        \subfloat[After relocating \TXb]{
          \includegraphics[width=\contourSize,trim=10 10 10 10,clip]{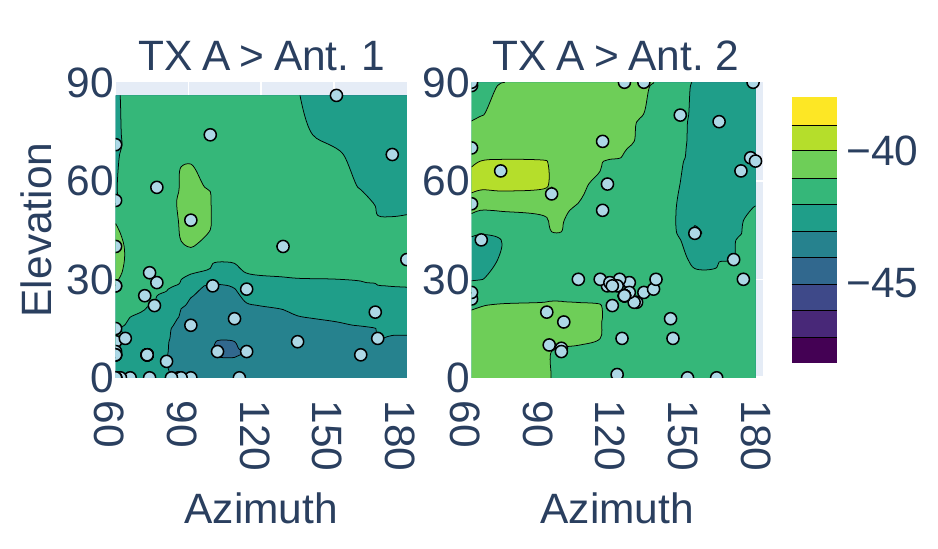}
          \includegraphics[width=\contourSize,trim=10 10 10 10,clip]{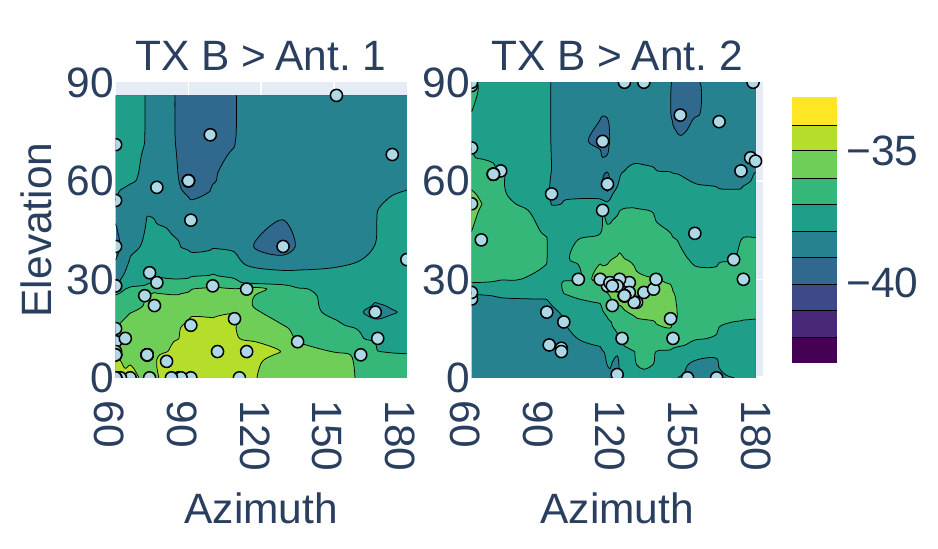}
        }
      \end{tabular}
    };
  \end{tikzpicture}
  \caption{
    Contour plots of the measured RSSI for the two mechanical antennas and the two TXs.
    White dots indicate the specific antenna orientations evaluated during the optimization process.
    The shifting optimal regions under different scenarios highlight the necessity of dynamic antenna tuning.
  }
  \label{fig:contour}
\end{figure}

To further illustrate the physical impact of the mechanical antennas, Fig.~\ref{fig:contour} visualizes the RSSI contour plots for the two independent antennas receiving signals from the two \TXpl.
The results reveal RSSI fluctuations of 3--6\,dB depending on the orientation, demonstrating the significant impact of antenna orientation on received signal strength.
While this substantial variation might suggest that a simpler, RSSI-based control mechanism could be sufficient, maximizing raw received power does not necessarily maximize the channel capacity or proportional fairness in a multi-\gls{STA} \gls{MIMO} environment.
Instead, the fact that the optimal angle regions shift significantly across the three scenarios reinforces the need for our capacity-based framework to adapt to the changes in the physical environment.

\section{Conclusion}
This paper proposed a mechanical Wi-Fi antenna device capable of autonomously tuning its 3D orientation to optimize communication performance for multiple stations.
To handle dynamic changes in active station combinations, as well as occasional environmental shifts, we developed an asynchronous framework equipped with state-specific optimizers and a capacity-based environment change detection mechanism.
Our experimental evaluations demonstrated that the proposed framework successfully adapts the antenna orientation during operation, improving proportional fairness among active transmitters.
Furthermore, we showed that the system robustly recovers from environmental blockages and station relocation.

Future work includes developing more computationally efficient optimization methods applicable to larger-scale systems with more stations and antennas, addressing highly dynamic scenarios such as mobile stations, and extending the framework toward \gls{ISAC} applications by jointly exploiting wireless measurements and mechanical antenna reconfiguration.

\bibliographystyle{IEEEtran}
\bibliography{IEEEabrv,myabrv,main}

\end{document}